\documentclass[reprint, superscriptaddress, amsmath, amssymb, aps]{revtex4-2}

\usepackage{graphicx}
\usepackage{bm}
\usepackage{braket}
\usepackage{qcircuit}
\usepackage{hyperref}
\usepackage{color}
\usepackage{physics}
\usepackage{amsmath}
\usepackage{amssymb}
\usepackage{float}
\usepackage{booktabs}

\begin{document}

\title{Feedback-Based Quantum Control for Safe and Synergistic Drug Combination Design}

\author{Mai Nguyen Phuong Nhi}
\affiliation{University of Science, Vietnam National University, Ho Chi Minh City 700000, Vietnam}
\affiliation{Vietnam National University, Ho Chi Minh City 700000, Vietnam}

\author{Lan Nguyen Tran}
\affiliation{University of Science, Vietnam National University, Ho Chi Minh City 700000, Vietnam}
\affiliation{Vietnam National University, Ho Chi Minh City 700000, Vietnam}

\author{Le Bin Ho}
\affiliation{Frontier Research Institute for Interdisciplinary Sciences, Tohoku University, Sendai 980-8578, Japan}
\affiliation{Department of Applied Physics, Graduate School of Engineering, Tohoku University, Sendai 980-8579, Japan}

\date{\today}

\begin{abstract}
Drug-drug interactions (DDIs) strongly affect the safety and efficacy of combination therapies. Despite the availability of large DDI databases, selecting optimal multi-drug combinations that balance safety, therapeutic benefit, and regimen size remains a challenging combinatorial optimization problem.
Here, we present a quantum-control-based framework for DDI-aware drug combination optimization, in which known harmful and synergistic interactions are encoded into Ising Hamiltonians as penalties and rewards, respectively. The optimization is performed using the feedback-based quantum algorithm FALQON, a gradient-free variational approach.
We study two clinically motivated tasks: the Maximum Safe Subset problem and the Synergy-Constrained Optimization problem. Numerical simulations using interaction data from Drugs.com and SYNERGxDB demonstrate efficient convergence and high-quality solutions for clinically relevant drug sets, including COVID-19 case studies.
\end{abstract}

\maketitle

\section{Introduction}
Drug-drug interactions (DDIs) are a common cause of harmful side effects, treatment failure, and hospital admission in modern medicine \cite{Abdu2025,Li2024,Zheng2018,Bucsa2013}. These problems become more serious as patients are increasingly treated with multiple drugs (polypharmacy) at the same time, for example in cancer therapy, long-term treatment of chronic diseases, or infectious diseases \cite{NOVAK2022778,Schmitz2025,jcm12123960,https://doi.org/10.1002/cncr.34642}. As more drugs are combined, the number of possible interactions grows very quickly, and even one dangerous combination can put a patient at serious risk \cite{NOVAK2022778,jcm12123960}.
Well-known examples include severe bleeding caused by certain combinations of blood thinners~\cite{Ballestri2023,DONZE2013502}, muscle damage~\cite{biomedicines12050987,doi:10.1152/physrev.00002.2019}, and kidney failure~\cite{Dormuthf880} that can occur when some cholesterol-lowering drugs are taken together with specific antibiotics. 

On the other hand, many effective treatments intentionally rely on positive drug interactions~\cite{DUARTE2022100110,SINGH2020221,SHYR20212367,CALZETTA20241159}. Combination therapies are widely used in HIV treatment \cite{SHYR20212367,Vanessa2011}, cancer chemotherapy \cite{SINGH2020221,OT16723}, and antibiotic regimens \cite{WORTHINGTON2013177}, where multiple drugs work together to improve effectiveness or prevent resistance. During the COVID-19 pandemic, several treatment strategies explored combinations of antiviral, anti-inflammatory, and supportive drugs, where both beneficial synergy and harmful interactions had to be carefully considered \cite{SHYR20212367,https://doi.org/10.1111/bcp.15089,BOBROWSKI2021873,Li2023}.

While substantial effort has been devoted to predicting DDIs and curating interaction databases \cite{Wang2024, Wang2025,doi:10.1073/pnas.1803294115}, an equally important downstream challenge remains: 
\emph{selecting safe and effective multi-drug regimens given existing DDI knowledge}. 
In clinical practice, interaction data are typically treated as fixed input, and the main difficulty lies in exploring the rapidly growing space of possible drug combinations while simultaneously avoiding harmful interactions and leveraging therapeutic synergy. 
This combinatorial complexity, together with the high cost of experimental and clinical validation, motivates the development of advanced optimization approaches, including quantum-assisted and quantum-inspired methods, for systematic drug combination design.

From a computational perspective, selecting an optimal subset of drugs from a library of size $n$ can be naturally formulated as a graph problem, where nodes represent drugs and edges encode harmful or synergistic interactions \cite{FILLINGER2025104345,LUO2024109148,10.1093,Wang2025,Chen2025,Wang2024,doi:10.1073/pnas.1803294115}. Finding the largest harm-free combination corresponds to a maximum independent-set problem on the harmful subgraph, while identifying a high-synergy, size-constrained combination with toxicity penalties leads to more general quadratic unconstrained binary optimization (QUBO) formulations \cite{glover2019tutorialformulatingusingqubo}. These problems are NP-hard and quickly become intractable for classical exhaustive search as $n$ grows, even when heuristic or approximate methods are employed \cite{PhysRevLett.127.120502}.

Quantum optimization algorithms provide a potential route to addressing such combinatorial explosions. The quantum approximate optimization algorithm (QAOA) has emerged as a widely studied framework for encoding discrete optimization problems into parameterized quantum circuits \cite{farhi2014quantumapproximateoptimizationalgorithm}. However, QAOA suffers from several limitations that are particularly severe on noisy intermediate-scale quantum (NISQ) devices: circuit depth grows with the number of layers, classical parameter optimization can be slow and prone to barren plateaus \cite{McClean2018}, and constrained problems often require delicate tuning of penalty terms. These challenges motivate the exploration of alternative quantum strategies that can handle complex constraint structures without incurring large classical overhead.

In this work, we introduce a quantum-control-based framework for \emph{DDI-aware drug combination optimization} based on the feedback-based quantum control algorithm (FALQON) \cite{PhysRevLett.129.250502} and its imaginary-time extension (ITE-FALQON) \cite{thanh2025}. 
FALQON eliminates the classical outer-loop optimization of QAOA by employing an intrinsic feedback law that updates a single control field using instantaneous quantum expectation values. 
The system evolves under a time-dependent Hamiltonian of the form $\mathcal{H}(t) = \mathcal{H}_{p} + \beta(t)\mathcal{H}_{d}$, where $\mathcal{H}_{p}$ encodes the DDI-aware objective and $\mathcal{H}_{d}$ is a simple driver Hamiltonian. 
This feedback mechanism enforces a monotonic decrease of the objective energy, effectively implementing quantum gradient descent without explicit gradient evaluation or multi-parameter tuning.

Within this framework, we formulate two complementary DDI optimization tasks. The first is the \emph{Maximum Safe Subset} (MSS) problem, which seeks the largest set of drugs that contains no harmful interactions and can be mapped to an Ising Hamiltonian with purely penalizing couplings. The second is the \emph{Synergy-Constrained Optimization} (SCO) problem, which augments the MSS model by incorporating synergy rewards and a cardinality constraint, thereby capturing the trade-off between efficacy, toxicity, and regimen size. Both objectives are encoded as Ising Hamiltonians derived from clinically documented interaction data obtained from sources such as Drugs.com and SYNERGxDB, with interaction strengths normalized to reflect relative severity or synergy.

We demonstrate our approach on a six-drug network consisting of Ritonavir, Everolimus, Cabazitaxel, Metformin, Erlotinib, and Topotecan. These drugs span distinct pharmacological classes and exhibit a complex structure of both harmful and synergistic interactions. We show that ITE-FALQON rapidly converges to low-energy configurations of the MSS and SCO Hamiltonians, outperforming standard FALQON, and reliably identifies clinically meaningful drug subsets. We further apply the method to a COVID-19 case study involving nine drug candidates. Our results highlight the advantages of feedback-based quantum control for constrained biomedical optimization problems. More broadly, they suggest that FALQON and related control-based algorithms offer a promising route toward quantum-assisted decision support for drug combination design.

The paper is organized as follows. In Sec.~\ref{sec2}, we introduce the theoretical framework and define two complementary DDI optimization tasks: MSS and SCO. Section~\ref{sec3} provides a brief overview of the optimization methods, including FALQON and ITE-FALQON. Numerical implementation details and benchmark results are presented in Sec.~\ref{sec4}. In Sec.~\ref{sec5}, we apply our approach to a COVID-19 case study. Finally, Secs.~\ref{sec6} and~\ref{sec7} are devoted to discussion and conclusions, respectively.

\section{Theoretical Framework}\label{sec2}
In this section, we introduce the mathematical framework used to encode DDI data into an Ising Hamiltonian suitable for quantum optimization. We begin with a drug library
\begin{align}
    \mathcal{D} = \{d_1, d_2, \ldots, d_n\},
\end{align}
where each drug $d_i$ is represented by a binary decision variable
\(
x_i \in \{0,1\},
\)
with $x_i = 1$ indicating that drug $i$ is included in the proposed combination and $x_i=0$ otherwise. A candidate multi-drug combination is represented by the bitstring $\bm{x} = (x_1, \ldots, x_n)$.

\begin{figure}[t] 
    \centering
\includegraphics[width=\columnwidth]{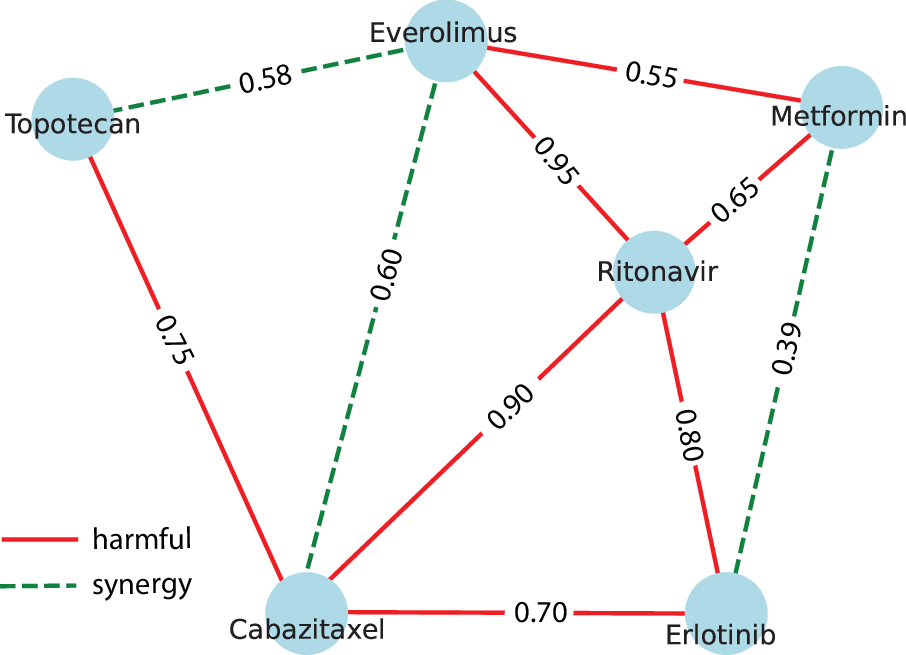} 
    \caption{Interaction graph of the six-drug network. Nodes represent individual drugs, while edges indicate pairwise interactions. Red edges denote harmful interactions and green edges denote synergistic interactions. Numbers on the edges indicate the corresponding interaction weights.}
    \label{fig1}
\end{figure}

The DDI data are encoded in a weighted graph
\begin{align}
    G = (\mathcal{D}, \mathcal{E}_{\mathrm{harm}} \cup \mathcal{E}_{\mathrm{syn}}),
\end{align}
where vertices represent drugs and edges denote either harmful (belong to a subset \(\mathcal{E}_{\mathrm{harm}}\)) or synergistic (belong to a subset \(\mathcal{E}_{\mathrm{syn}}\)) interactions. Each edge $(i,j)$ is assigned a weight that quantifies the clinical severity of a harmful interaction or the strength of a synergistic effect. These values are extracted from established pharmacological resources, including Drugs.com and SYNERYxDB.ca and normalized to the interval $[0,1]$. In Fig.~\ref{fig1}, we illustrate a graph of 6 drugs as summarized in Tab.~\ref{tab1}. 

\begin{table}[h!]
\centering
\caption{DDI dataset used in the simulation (weights inferred from clinical severity).}
\begin{tabular}{c c c c}
\toprule
${\rm drug}_i$   & ${\rm drug}_j$   & kind     & weight $w_{ij}^{\rm kind}$ \\
\midrule
Ritonavir        & Everolimus       & harm     & 0.95 \\
Ritonavir        & Cabazitaxel      & harm     & 0.90 \\
Metformin        & Ritonavir        & harm     & 0.65 \\
Metformin        & Everolimus       & harm     & 0.55 \\
Ritonavir        & Erlotinib        & harm     & 0.80 \\
Topotecan        & Cabazitaxel      & harm     & 0.75 \\
Erlotinib        & Cabazitaxel      & harm     & 0.70 \\
Everolimus       & Cabazitaxel      & synergy  & 0.60 \\
Everolimus       & Topotecan        & synergy  & 0.58 \\
Erlotinib        & Metformin        & synergy  & 0.39 \\
\bottomrule
\end{tabular}\label{tab1}
\end{table}

The goal of the optimization is to choose a group of drugs that is safe to use together and, when possible, provides useful therapeutic benefits. To do this, we build an objective function that accounts for harmful interactions, possible synergy between drugs, and the preferred number of drugs in the combination. We introduce two models below that describe how these requirements are encoded.

\subsection{Maximum Safe Subset (MSS)}
The first task is the \emph{Maximum Safe Subset} (MSS) problem. Here, the objective is to choose the largest possible set of drugs that contains no harmful interactions. In other words, we seek the largest drug combination in which all drugs can be safely used together.

To encode this objective, we assign a reward for selecting each drug and impose a penalty for selecting any harmful pair. The loss function is
\begin{equation}
\mathcal{L}_{\text{MSS}}(\bm x) 
= 
- \sum_{i=1}^{n} x_i 
\;+\;
\alpha \sum_{(i,j) \in \mathcal{E}_{\text{harm}}} 
w_{ij}^{\text{harm}} x_i x_j,
\label{eq:MSS_loss}
\end{equation}
where $\alpha>0$ is a penalty coefficient controlling the severity of constraint enforcement and \(w_{ij}^{\rm harm}\) are harmful weights.
The first term encourages selecting as many drugs as possible, while the second term penalizes the inclusion of any harmful pair. If $\alpha$ is sufficiently large, the optimal solution will exclude all harmful pairs entirely.

To convert Eq.~\eqref{eq:MSS_loss} into an Ising Hamiltonian, we map
\[
x_i = \frac{1 - z_i}{2},
\]
where $z_i \in \{-1,+1\}$ are Pauli-$Z$ eigenvalues. Substituting this relation into Eq.~\eqref{eq:MSS_loss} yields an Ising Hamiltonian
\begin{equation}
\mathcal{H}_{\mathrm{MSS}}
=
c^{\mathrm{MSS}}
+ \sum_{i} h_i^\mathrm{MSS} Z_i
+ \sum_{i<j} J_{ij}^{\mathrm{MSS}} Z_i Z_j,
\end{equation}
where \(Z_i\) is the Pauli-\(Z\) operator acting on qubit \(i\), and the coefficients determined explicitly by
\begin{align}
c^{\text{MSS}} 
&= -\frac{n}{2} 
 + \frac{\alpha}{4} \sum_{(i,j)\in\mathcal{E}_{\mathrm{harm}}}
   w_{ij}^{\mathrm{harm}}, \\
h_i^\mathrm{MSS}
&= \frac{1}{2}
 - \frac{\alpha}{4}
   \sum_{(i,j)\in\mathcal{E}_{\mathrm{harm}}}
   w_{ij}^{\mathrm{harm}}, \\
J_{ij}^{\mathrm{MSS}}
&= \frac{\alpha}{4} w_{ij}^{\mathrm{harm}}.
\end{align}
Thus, harmful interactions generate positive Ising couplings $J_{ij}^{\rm MSS}>0$, raising the energy of configurations where both $z_i=z_j=-1$ (i.e., $x_i=x_j=1$). The ground state of this Hamiltonian corresponds exactly to the largest safe drug subset.

\subsection{Synergy-Constrained Optimization (SCO)}
While the MSS formulation focuses exclusively on safety by eliminating harmful drug pairs, many therapeutic applications demand a more nuanced balance between efficacy and toxicity~\cite{CHOU2006621}. In clinical practice, desirable drug combinations simultaneously exploit beneficial synergy, avoid adverse interactions, and meet a prescribed therapeutic size. To accommodate these competing requirements, we introduce the \emph{Synergy-Constrained Optimization} (SCO) model, which extends the MSS formalism by incorporating synergistic rewards and a cardinality constraint.

In SCO, the objective function contains three contributions. First, synergistic interactions decrease the loss, favoring the co-selection of pharmacologically compatible drug pairs. Second, harmful interactions increase the loss through a toxicity penalty weighted by a parameter $\gamma$, which allows us to modulate the strictness of harm suppression. Third, a quadratic regularization term enforces a target combination size $K$, ensuring that the resulting drug set satisfies application-specific design criteria. These considerations lead to the loss function
\begin{align}
\mathcal{L}_{\text{SCO}}(\bm x)
&=
-
\sum_{(i,j)\in \mathcal{E}_{\mathrm{syn}}}
w_{ij}^{\mathrm{syn}} x_i x_j
+
\gamma
\sum_{(i,j)\in \mathcal{E}_{\mathrm{harm}}}
w_{ij}^{\mathrm{harm}} x_i x_j
\nonumber\\
&\quad
+
\mu
\left(
\sum_{i=1}^{n} x_i - K
\right)^2,
\label{eq:SCO_loss}
\end{align}
where $\gamma$ tunes the harmful penalty, $\mu$ controls the strength of the size constraint, and $K$ denotes the desired number of selected drugs. The quadratic term introduces both local fields and effective all-to-all couplings once expanded, reflecting the global nature of the cardinality requirement.

Substituting the standard Ising mapping $x_i = (1 - z_i)/2$ into Eq.~\eqref{eq:SCO_loss} yields an Ising Hamiltonian as
\begin{align}
    \mathcal{H}_{\text{SCO}}
=
c^{\text{SCO}}
+
\sum_i h_i^{\text{SCO}} Z_i
+
\sum_{i<j} J_{ij}^{\text{SCO}} Z_i Z_j,
\end{align}
where 
\begin{align}
\notag c^{\rm SCO} &= -\frac{1}{4} \sum_{(i,j) \in \mathcal{E}_{\rm syn}} w_{ij}^{\rm syn} \\
&+ \frac{\gamma}{4} \sum_{(i,j) \in \mathcal{E}_{\rm harm}} w_{ij}^{\rm harm} + \mu \left(\frac{n}{2} - K\right)^2 + \mu \frac{n}{4}, \\
h_i^{\text{SCO}} &= \frac{1}{4} \sum_{j \in \mathcal{E}_{\rm syn}} w_{ij}^{\rm syn} - \frac{\gamma}{4} \sum_{j \in \mathcal{E}_{\rm harm}} w_{ij}^{\rm harm} - \mu \left(\frac{n}{2} - K\right), \\
J_{ij}^{\text{SCO}} &= -\frac{1}{4} w_{ij}^{\rm syn} + \frac{\gamma}{4} w_{ij}^{\rm harm} + \frac{\mu}{2}.
\end{align}

The sign and magnitude of the couplings have clear pharmacological meaning: synergistic pairs generate negative couplings, which energetically favor their co-selection, harmful pairs generate positive couplings, which penalize configurations containing both drugs, and the cardinality term induces global correlations that steer the system toward the prescribed combination size. Consequently, the ground state of $\mathcal{H}_{\text{SCO}}$ encodes the optimal clinically constrained drug set, simultaneously balancing efficacy, safety, and size.

\section{Quantum Optimization Methods}\label{sec3}
To find the ground energy and ground state (bitstring solution) of $\mathcal{H}_{\rm MSS}$ and $\mathcal{H}_{\rm SCO}$, we apply 
the feedback-based quantum control algorithm (FALQON)~\cite{PhysRevLett.129.250502}
and a recent developed method imaginary-time evolution enhanced FALQON (ITE-FALQON)~\cite{thanh2025}.

\subsection{FALQON}
Unlike hybrid variational methods such as QAOA, which rely on repeated circuit executions combined with classical parameter optimization, FALQON determines its control parameters directly from the quantum dynamics. This eliminates the need for classical training and makes the method particularly effective for constrained combinatorial problems with highly nonconvex energy landscapes.

The quantum state evolves under a time-dependent Hamiltonian
\begin{align}
    \mathcal{H}(t) = \mathcal{H}_p + \beta(t) \mathcal{H}_d,
\end{align}
where \(\mathcal{H}_p\) is the problem Hamiltonian, i.e., $\mathcal{H}_{\rm MSS}, \mathcal{H}_{\rm SCO}$, and \(\mathcal{H}_d=\sum_i X_i\) is a driver Hamiltonian that induces transitions between computational basis states. All pharmacological information is contained in \(\mathcal{H}_p\); therefore, its ground state corresponds to the optimal drug combination.

The feedback mechanism adjusts the control field \(\beta(t) = -i\langle\psi(t)| [\mathcal{H}_{d},\mathcal{H}_{p}] |\psi(t)\rangle\) such that the expectation value
\begin{align}
    C(t)=\langle \psi(t)|\mathcal{H}_p|\psi(t)\rangle,
\end{align}
decreases monotonically during the evolution. The update rule is determined by the instantaneous commutator between \(\mathcal{H}_d\) and \(\mathcal{H}_p\), which identifies the direction of energy descent. In this way, FALQON implements a quantum analogue of gradient descent without explicitly computing gradients or optimizing multiple variational parameters.

In practice, the evolution is discretized into a sequence of alternating applications of \(\mathcal{H}_p\) and \(\mathcal{H}_d\), with the control field updated at each step through the feedback rule. The circuit depth therefore grows linearly with the number of time steps, making the algorithm robust to noise and suitable for near-term quantum hardware. In the DDI setting, this dynamical process naturally suppresses unsafe drug combinations while amplifying configurations that satisfy the MSS or SCO objectives.

\subsection{ITE-FALQON}

To further accelerate convergence toward the ground state and suppress residual excited-state components, we consider an extension of FALQON based on imaginary-time evolution or ITE-FALQON~\cite{thanh2025}. In imaginary time, the state propagation is governed by the non-unitary transformation
\[
|\psi(\tau+\Delta\tau)\rangle \propto e^{-\Delta\tau \mathcal{H}_p} |\psi(\tau)\rangle,
\]
which acts as an energy filter.

By expanding the state in the energy eigenbasis of \(\mathcal{H}_p\), imaginary-time evolution exponentially suppresses higher-energy components relative to the ground state. As the imaginary time \(\tau\) increases, the state is progressively projected onto the lowest-energy subspace, thereby enhancing ground-state fidelity.

Within the ITE-FALQON framework, this energy-filtering mechanism is combined with the feedback-controlled dynamics of FALQON. The resulting evolution guides the system more directly toward the true ground state of the $\mathcal{H}_{\rm MSS}$ and $\mathcal{H}_{\rm SCO}$, improving convergence speed and stability, particularly in cases where the energy spectrum is dense or near-degenerate~\cite{thanh2025}.

\section{Implementation Details}\label{sec4}

\subsection{Model setup}
We construct a six-drug interaction network from clinically documented data, as shown in Fig.~\ref{fig1} and Tab.~\ref{tab1}. The drug set includes Ritonavir, Everolimus, Cabazitaxel, Metformin, Erlotinib, and Topotecan, which covers antiviral, oncological, and metabolic therapies and exhibits a range of harmful and synergistic interactions. This combination provides a compact yet representative testbed for DDI optimization. 
The interaction graph contains weighted harmful edges, which introduce energetic penalties, and synergistic edges, which lower the energy of favorable combinations. This graph fully determines the Ising Hamiltonians used in the FALQON and ITE-FALQON simulations.

All numerical simulations are performed using the Qiskit statevector backend. 
Each drug is encoded by a single qubit, and the corresponding Ising Hamiltonians are constructed from standard Pauli-$Z$ and $ZZ$ operators according to $\mathcal{H}_{\rm MSS}$ and $\mathcal{H}_{\rm SCO}$. 
The mixer Hamiltonian consists of a sum of single-qubit Pauli-$X$ operators.

The quantum evolution is discretized into $T$ time steps. 
At each step, the control field $\beta(t)$ is updated using the FALQON feedback rule. 
For ITE-FALQON, an imaginary-time evolution step is applied after every two FALQON updates. 
The imaginary-time step size $\Delta\tau$ is chosen separately for each model to account for differences in energy scales. 
For the MSS problem, we use $\Delta\tau = 0.1$, while the SCO problem requires a smaller step size, $\Delta\tau = 0.01$, to ensure numerical stability.
This smaller $\Delta\tau$ is necessary because the SCO Hamiltonian includes a quadratic cardinality constraint, $\mu\left(\sum_i x_i - K\right)^2$, which leads to larger energy eigenvalues and all-to-all couplings. 

After the final evolution, measurements yield a probability distribution over all bitstrings, where the most probable low-energy states correspond to candidate drug combinations satisfying the MSS or SCO criteria.

\subsection{MSS Results}

\begin{figure}[t]
    \centering
    \includegraphics[width=\columnwidth]{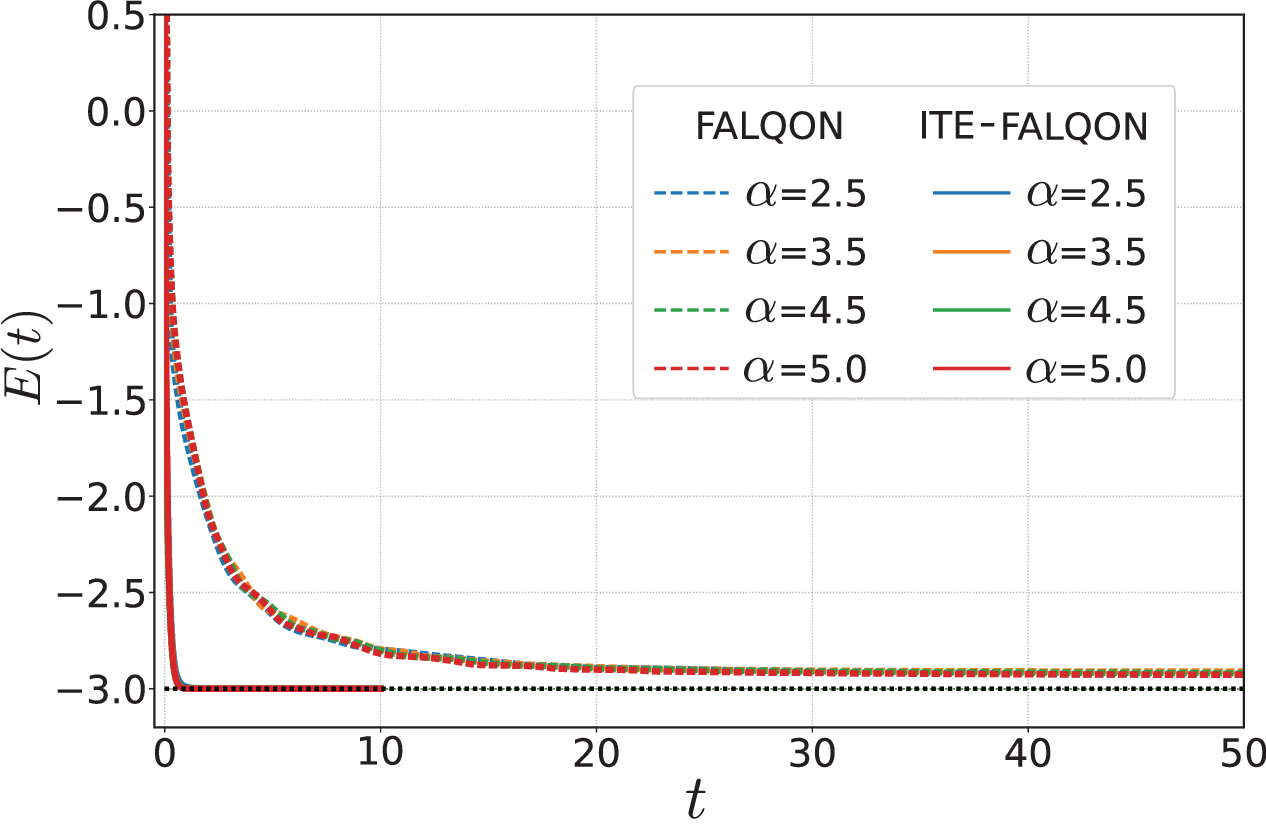}
    \caption{Energy evolution obtained using FALQON and ITE-FALQON for different penalty coefficients \(\alpha\). While FALQON reliably identifies the correct optimal bitstring, its final energy remains slightly above the exact ground-state energy due to the purely unitary nature of the dynamics, which does not explicitly suppress excited-state components. In contrast, ITE-FALQON rapidly reaches the ground-state energy by effectively filtering out higher-energy contributions through imaginary-time evolution. This comparison highlights the complementary roles of feedback-based control and non-unitary energy filtering in accelerating convergence.}
    \label{fig2}
\end{figure}

We first investigate the MSS problem by applying FALQON and ITE-FALQON to the Hamiltonian \(\mathcal{H}_{\mathrm{MSS}}\). Figure~\ref{fig2} shows the time evolution of the energy
\(
E(t)=\langle\psi(t)|\mathcal{H}_{\mathrm{MSS}}|\psi(t)\rangle
\)
for several values of the penalty parameter \(\alpha\), which controls the energetic cost of violating harmful-interaction constraints.

For all values of \(\alpha\), FALQON drives the system toward the correct low-energy subspace and reliably identifies the optimal safe configuration. However, because the evolution is purely unitary, the final energy saturates slightly above the exact ground-state value. This residual offset reflects the presence of excited-state components that are not explicitly suppressed under unitary dynamics.

In contrast, ITE-FALQON converges rapidly and reaches the exact ground-state energy for all tested values of \(\alpha\). The imaginary-time component effectively filters out higher-energy contributions, leading to a sharp projection onto the ground state and significantly faster convergence.

The role of the penalty parameter \(\alpha\) is also evident. Increasing \(\alpha\) amplifies the energetic separation between valid (harm-free) configurations and invalid ones, thereby enlarging the spectral gap of \(\mathcal{H}_{\mathrm{MSS}}\). This enhanced separation accelerates convergence in both algorithms, with the effect being particularly pronounced for ITE-FALQON, which exploits the widened gap through exponential suppression of excited states.

These results demonstrate that while FALQON is sufficient to identify the correct MSS solution, the inclusion of imaginary-time evolution substantially improves convergence speed and accuracy by enforcing explicit energy filtering.

\begin{figure}[t]
    \centering
    \includegraphics[width=\columnwidth]{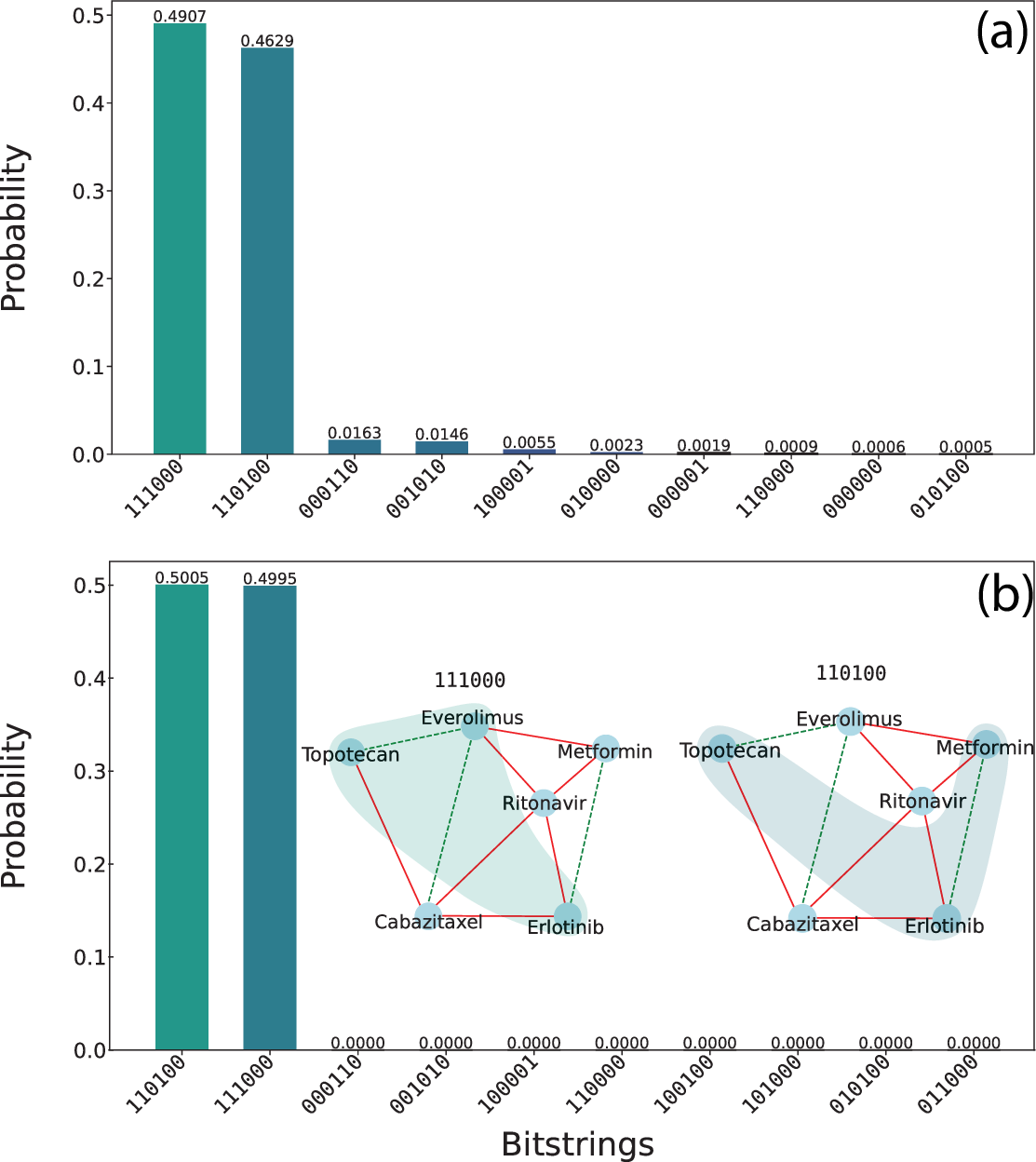} 
    \caption{Final probability distributions and dominant MSS solutions. (a) FALQON concentrates probability on two harm-free bitstrings, \(111000\) and \(110100\), while suppressing unsafe configurations. (b) ITE-FALQON further sharpens this distribution, projecting almost entirely onto the same two degenerate MSS solutions. The shaded graphs depict the corresponding drug subsets.}
    \label{fig3}
\end{figure}

Figure~\ref{fig3}(a) shows the measurement probabilities after the FALQON evolution. As expected for the MSS objective, significant probability accumulates only on configurations that contain no harmful interactions. The distribution is strongly dominated by two bitstrings, \(111000\) and \(110100\), with probabilities of approximately \(0.49\) and \(0.46\), respectively. All remaining bitstrings appear with probabilities below \(2\times10^{-2}\), indicating that unsafe configurations are effectively suppressed.

Importantly, the identity of these dominant bitstrings is independent of the specific value of the penalty coefficient \(\alpha\), provided that \(\alpha\) is large enough to energetically separate unsafe states. Increasing \(\alpha\) therefore does not change which configurations are optimal, but instead reinforces the energetic separation between feasible and infeasible subsets. In Fig.~\ref{fig3}, we use $\alpha = 5$. The coexistence of two dominant solutions reflects the fact that the MSS problem admits multiple maximum safe subsets with equal cardinality and nearly degenerate energies.

\begin{figure*}[t]
    \centering
    \includegraphics[width=\linewidth]{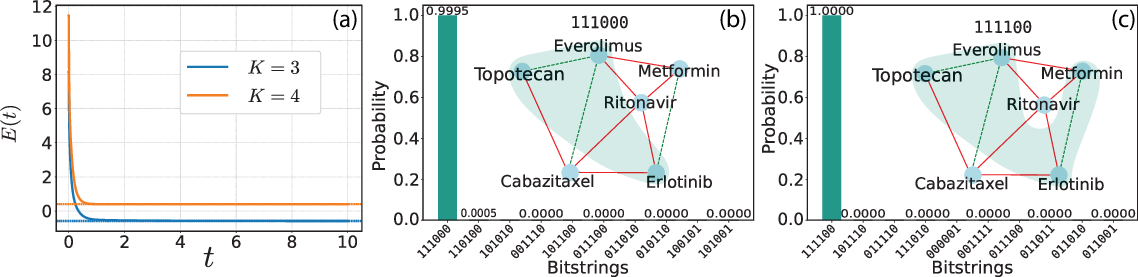} 
    \caption{SCO results obtained using ITE-FALQON. (a) Energy convergence for target cardinalities \(K=3\) and \(K=4\). (b, c) Final measurement probability distributions and corresponding interaction graphs for \(K=3\) and \(K=4\), respectively. In both cases, the dynamics concentrate almost entirely on a single dominant bitstring, indicating a unique optimal drug combination. The inset graphs visualize the selected subsets and their internal interactions, illustrating the balance between synergistic benefit and interaction risk as the cardinality constraint is increased.}
    \label{fig4}
\end{figure*}

Figure~\ref{fig3}(b) shows the corresponding results obtained using ITE-FALQON. In this case, the probability distribution becomes even more sharply concentrated: essentially all probability mass collapses onto the same two bitstrings, \(111000\) and \(110100\), with nearly equal weights. All other configurations are suppressed to negligible probabilities. This behavior is a direct consequence of the imaginary-time component, which exponentially filters out excited-state contributions and projects the system more efficiently onto the ground-state subspace \cite{thanh2025}.

The network visualizations in Fig.~\ref{fig3}(b) illustrate the drug subsets represented by these two bitstrings. Shaded nodes indicate the selected drugs. Both configurations exclude all major harmful interactions, confirming that they correspond to valid maximum safe subsets. The near-equal probabilities and energies of the two solutions highlight the intrinsic degeneracy of the MSS problem and demonstrate that ITE-FALQON faithfully recovers the complete set of optimal safe regimens.

\subsection{SCO Results}

We now turn to the SCO problem, which extends the MSS formulation by incorporating both synergistic rewards and an explicit cardinality constraint. The objective is to identify drug combinations of fixed size \(K\) that maximize therapeutic synergy while suppressing harmful interactions. In the results shown here, we apply the ITE-FALQON algorithm and compare two target sizes, \(K=3\) and \(K=4\), why we fix
$\gamma=2.5$ and $\mu=5$.

Figure~\ref{fig4} summarizes the SCO performance. Fig.~\ref{fig4}(a) shows the time evolution of the energy \(E(t)=\langle\psi(t)|\mathcal{H}_{\mathrm{SCO}}|\psi(t)\rangle\). In both cases of \(K\), the energy rapidly decreases and stabilizes within a short optimization time, indicating efficient suppression of excited-state contributions. The difference in the final energy reflects the imposed cardinality constraint: increasing \(K\) shifts the optimal energy upward due to the larger number of selected drugs and the associated interaction terms.

Figure~\ref{fig4} (b, c) show the final measurement probability distributions for \(K=3\) and \(K=4\), respectively. In both cases, the dynamics concentrate almost entirely on a single bitstring, demonstrating that the optimization uniquely selects a dominant optimal solution. The corresponding interaction graphs highlight that the selected drug subsets simultaneously satisfy the size constraint, avoid harmful internal interactions, and preferentially include synergistic pairs.

These results show that the SCO formulation reliably extracts clinically interpretable drug combinations. The dominant bitstrings correspond to subsets that are internally consistent with known DDI structure, while the changes observed between \(K=3\) and \(K=4\) directly reflect the underlying pharmacological constraints encoded in the interaction network.

\section{COVID-19 Case Study}\label{sec5}
During the early stages of the COVID-19 pandemic, many drugs were tested in clinical trials to find effective combinations against SARS-CoV-2. Because a large number of possible drug combinations were proposed, identifying the best subsets quickly became a challenging combinatorial optimization problem. Motivated by this challenge, we apply the ITE-FALQON method to a panel of nine drugs to examine whether clinically effective drug subsets can be recovered by minimizing a Hamiltonian constructed from pairwise drug interactions.

The drugs are grouped into three functional classes: viral replication inhibitors, entry and post-translational inhibitors, and immune response regulators. The diversity of these mechanisms leads to both cooperative and conflicting interactions, creating a complex optimization landscape that is well suited for benchmarking the ITE-FALQON framework. Table~\ref{tab2} summarizes the pharmacological mechanisms and abbreviations used for each drug in this study.

\begin{table*}[t]
\centering
\caption{Classification and mechanism of action for 9 drugs included in the COVID-19 repurposing case study. Abbreviations are used in the interaction graphs and Hamiltonian encoding.}
\label{tab2}
\begin{tabular}{l c l}
\toprule
\textbf{Drug Name} & \textbf{Abbreviation} & \textbf{Primary Mechanism / Pharmacological Class} \\
\midrule
\multicolumn{3}{l}{\textit{Viral Replication Inhibitors (Antivirals)}} \\
Remdesivir & RDV & Nucleoside analog; RdRp inhibitor \cite{Rahmah2022-tn, Brady2024-if} \\
Molnupiravir & MOV & Ribonucleoside analog; induces lethal mutagenesis \cite{Rahmah2022-tn, Brady2024-if} \\
Favipiravir & FPV & Purine analog; selective RdRp inhibitor \cite{Rahmah2022-tn, Brady2024-if} \\
Ribavirin & RBV & Guanosine analog; interferes with RNA synthesis \cite{Rahmah2022-tn, Brady2024-if} \\
Nirmatrelvir/Ritonavir & PAX & SARS-CoV-2 main protease (Mpro/3CLpro) inhibitor \cite{Rahmah2022-tn, Brady2024-if} \\
Lopinavir/Ritonavir & LPV & HIV protease inhibitor (repurposed) \cite{Rahmah2022-tn, Brady2024-if} \\
\midrule
\multicolumn{3}{l}{\textit{Entry and Post-Translational Inhibitors}} \\
Nitazoxanide & NTZ & Broad-spectrum; inhibits viral entry/assembly \cite{Rahmah2022-tn, Brady2024-if} \\
Hydroxychloroquine & HCQ & Antimalarial; modulates endosomal pH \cite{Rahmah2022-tn, Brady2024-if} \\
\midrule
\multicolumn{3}{l}{\textit{Immune Response Regulation}} \\
Dexamethasone & DEX & Corticosteroid; anti-inflammatory/immunomodulator \cite{Rahmah2022-tn, Brady2024-if} \\
\bottomrule
\end{tabular}
\end{table*}

\begin{figure}[b] 
    \centering
\includegraphics[width=\columnwidth]{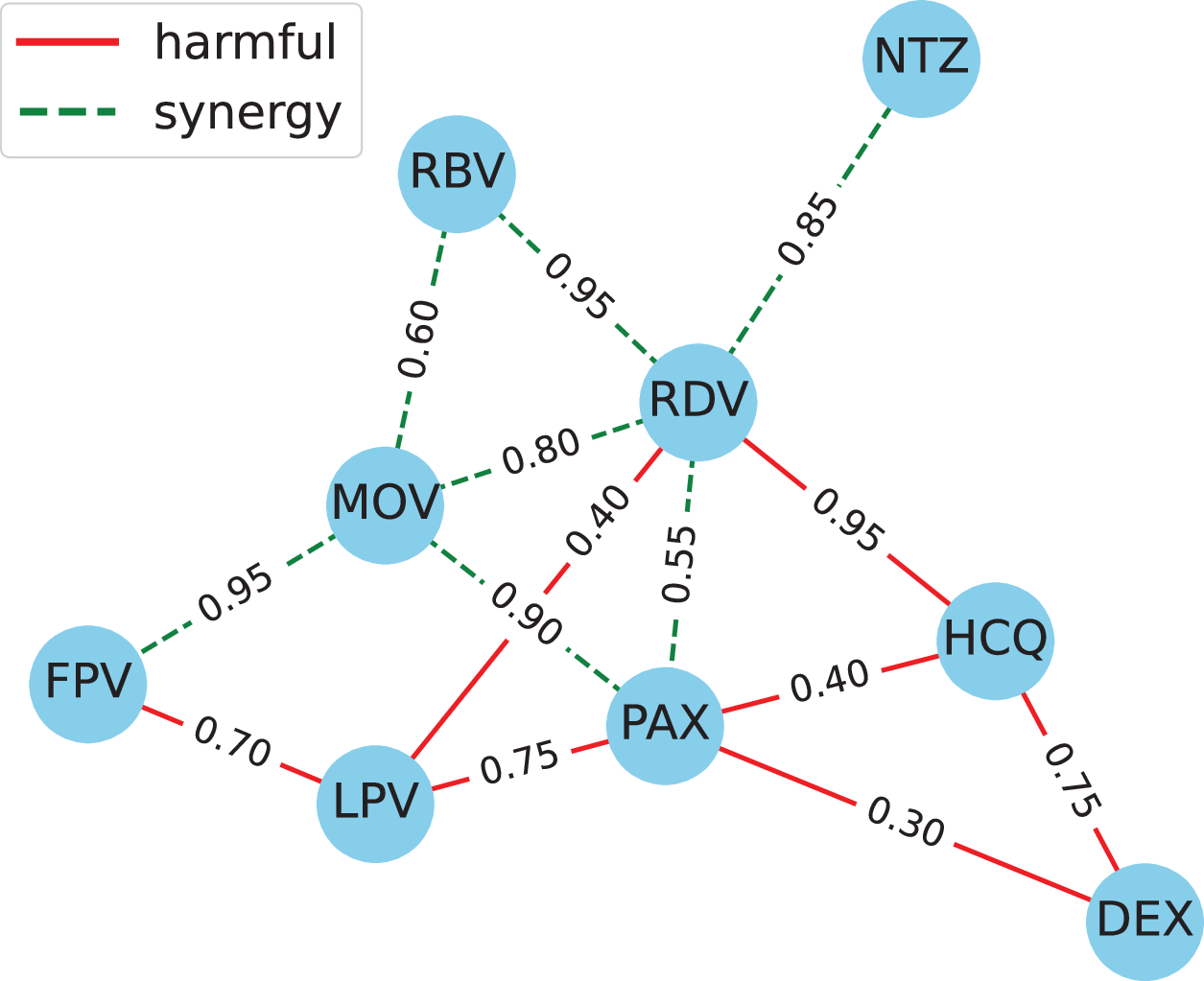} 
    \caption{Interaction graph of the 9-drug network.} 
    \label{fig5}
\end{figure}

Pairwise synergy and harmful interaction weights were obtained from the Liverpool COVID-19 Drug Interaction database and supporting pharmacological studies. These weights were normalized to the range $[0,1]$, where strongly complementary antiviral pairs were assigned high synergy, and harmful interactions were identified based on reported adverse effects (see Table~\ref{tab3}). For example, Remdesivir and Ribavirin show synergistic antiviral activity and are therefore assigned a high synergy weight, whereas the combination of Remdesivir and Hydroxychloroquine is known to reduce antiviral efficacy and is classified as harmful. The interaction graph is shown in Fig.~\ref{fig5}.

This procedure encodes the known drug interaction structure directly into the Hamiltonian while remaining robust to uncertainties in quantitative interaction strengths. Importantly, the optimization relies only on pairwise interactions, with no higher-order terms or training data introduced.

\begin{figure}[t]
    \centering
    \includegraphics[width=\columnwidth]{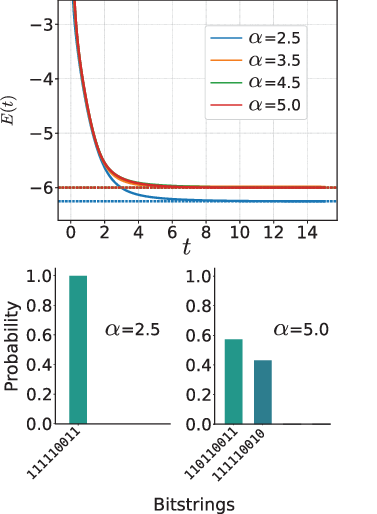} 
    \caption{(upper) Energy evolution obtained from the ITE-FALQON for different penalty coefficients \(\alpha\). (lower) Final probability distributions and dominant MSS solutions. (left) For relatively small values of $\alpha$ (e.g., $\alpha$=2.5), the optimal solutions is $\{$RDV, RBV, PAX, NTZ, MOV, FPV, DEX$\}$, which still contains harmful terms. (right) For $\alpha$ = 5, we get $\{$RDV, RBV, NTZ, MOV, FPV, DEX$\}$, which no harmful interaction at all.}
    \label{fig:6}
\end{figure}

\begin{table}[h!]
\centering
\caption{COVID-19 DDI dataset (weights from literature).}
\begin{tabular}{c c c c}
\toprule
${\rm drug}_i$   & ${\rm drug}_j$   & kind     & weight $w_{ij}^{\rm kind}$ \\
\midrule
RBV              & RDV              & synergy  & 0.95 \\
MOV              & PAX              & synergy  & 0.90 \\
MOV              & RDV              & synergy  & 0.80 \\
MOV              & RBV              & synergy  & 0.60 \\
PAX              & RDV              & synergy  & 0.55 \\
FPV              & MOV              & synergy  & 0.95 \\
NTZ              & RDV              & synergy  & 0.85 \\
HCQ              & RDV              & harm     & 0.95 \\
HCQ              & PAX              & harm     & 0.40 \\
LPV              & PAX              & harm     & 0.75 \\
LPV              & RDV              & harm     & 0.40 \\
DEX              & HCQ              & harm     & 0.75 \\
DEX              & PAX              & harm     & 0.30 \\
FPV              & LPV              & harm     & 0.70 \\
\bottomrule
\end{tabular}\label{tab3}
\end{table}

\subsection{MSS Results}
When applying the MSS Hamiltonian to this drug set, we observe a clear dependence of the low-energy solutions on the penalty factor \(\alpha\). For relatively small values of \(\alpha\) (e.g., \(\alpha\)=2.5), the optimal solutions is $\{$RDV, RBV, PAX, NTZ, MOV, FPV, DEX$\}$, which still contains harmful terms, such as the PAX-DEX pair because the energy of adding a new drug can outweigh the penalty imposed by a harmful interaction. As a result, the algorithm accepts a trade-off between maximizing the subset size and tolerating limited safety violations. 

As the penalty factor is increased, the tendency of energy landscape starts to change: all dominant low-energy configurations systematically exclude harmful pairs. For example, with \(\alpha\) = 5, we get $\{$RDV, RBV, NTZ, MOV, FPV, DEX$\}$, which no harmful interaction at all.
This behavior is consistent with the MSS objective, which identifies the largest subset of drugs that remains compliant with safety constraints. In conclusion, penalty factor \(\alpha\) plays an important role in reflecting different levels of clinical risk tolerance, and larger \(\alpha\) is better.

\subsection{SCO Results}
We next consider the SCO setting by imposing a specific constraint on the selected subset, fixing the number of drugs to $K=3$. Under this constraint, we obtain the most dominant bitstring corresponding to the drug triples $\{$RDV, RBV, MOV$\}$. 
Moreover, in vitro and animal model studies have shown that both combinations can enhance antiviral efficacy including synergistic suppression of SARS-CoV-2 replication. Compared to the MSS formulation, which favors selecting as many mutually non-harmful drugs as possible, the SCO formulation is more aligned with clinical practice by restricting the size of the optimal drug subset. In the context of actual COVID-19 treatment, increasing the number of co-administered drugs may increase the risk of toxicity due to higher-order polypharmacy effects, even in the absence of pairwise conflicts. 
This case study serves as a proof-of-principle that ITE-FALQON framework can recover clinically meaningful drug combinations within a large combinatorial search space using only pairwise interaction information encoded in a physical Hamiltonian. 

\begin{figure}[t]
    \centering 
    \includegraphics[width=\columnwidth]{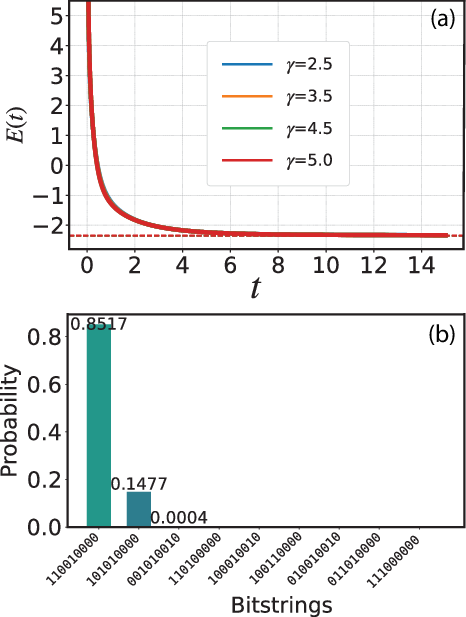} 
    \caption{(a) Energy evolution obtained from ITE-FALQON for different penalty coefficients \(\gamma\). (b) Under this constraint, we obtain the most dominant bitstring corresponding to the drug triples $\{$RDV, RBV, MOV$\}$.
    }
    \label{fig:7}
\end{figure}

\section{Discussion}\label{sec6}
The present results show that FALQON and ITE-FALQON can reliably identify safe or synergistic subsets in small DDI networks, and the structure of real pharmacological data suggests that the method can scale beyond the six-drug examples studied here. Large clinical DDI databases such as DrugBank or SYNERGxDB contain hundreds of drugs, but the corresponding interaction graphs are highly sparse: only a small fraction of drug pairs exhibit clinically documented interactions. This sparsity substantially reduces the number of nonzero couplings in the Ising Hamiltonian and therefore mitigates the growth of circuit complexity as the number of drugs increases.

A second feature of real DDI networks is their modularity. Harmful or synergistic interactions tend to cluster within drug classes, allowing the Hamiltonian to be organized into blocks that can be encoded efficiently. Together, sparsity and modularity imply that Hamiltonians derived from realistic DDI datasets remain tractable for statevector simulations and for near-term quantum hardware in the 50-100 qubit range, provided that appropriate preprocessing steps are applied.

From an algorithmic perspective, FALQON and its extend version offer clear advantages over variational approaches such as QAOA. Because the control field is updated through a closed-loop quantum expectation value rather than a classical optimization routine, FALQON avoids barren plateaus and does not require deep, parameter-heavy circuits. Its depth grows only linearly with the number of time steps, making the method robust to noise and well suited to constrained optimization problems like MSS and SCO, where large penalties induce steep energy landscapes.

These properties suggest that FALQON provides a promising route toward quantum-assisted analysis of clinically relevant DDI networks. Embedding real-world synergy and toxicity scores directly into the Hamiltonian is straightforward, and future work will focus on scaling to larger datasets and incorporating richer pharmacokinetic constraints.

\section{Conclusion}\label{sec7}
We have presented a quantum control-based framework for DDI-aware drug combination optimization using the FALQON and ITE-FALQON algorithms. 
By encoding clinically documented drug-drug interaction data into Ising Hamiltonians, our approach efficiently identifies safe and synergistic multi-drug combinations with reduced computational overhead. 
These results highlight the potential of feedback-based quantum control algorithms as practical tools for constrained combinatorial optimization in biomedical decision support and rational multi-drug therapy design.

\acknowledgments
This paper is supported by JSPS
KAKENHI Grant Number 23K13025, and the Tohoku Initiative for Fostering Global Researchers for Interdisciplinary Sciences (TI-FRIS) of MEXT's Strategic Professional Development Program for Young Researchers. L.N.T is funded by National Foundation for Science and Technology Development (NAFOSTED) Grant Number 103.01-2024.06.
M.N.P.N acknowledges helpful discussions with Nguyen Van Long Thanh on the FALQON method.

\section*{Data availability} 
The data that support the findings of this article are openly
available \cite{nhi2026}.

\bibliography{refs}
\end{document}